\documentstyle{article}
\newcount\Mac  \Mac=0 
\newcommand{\comm}[1]{}

\newcommand{\rt}{\tilde{r}}
\newcommand{\Ut}{\tilde{U}}

\newcommand{\At}{\tilde{A}}
\newcommand{\St}{\tilde{S}}

\newcommand{\phit}{\tilde{\phi}}

\newcommand{\lx}{\lambda}

\newcommand{\be}{\begin{equation}}
\newcommand{\ee}{\end{equation}}
\newcommand{\een}{\end{subequations}}
\newcommand{\ben}{\begin{subequations}}
\newcommand{\beq}{\begin{eqnarray}}
\newcommand{\eeq}{\end{eqnarray}}

\def \lta {\mathrel{\vcenter
     {\hbox{$<$}\nointerlineskip\hbox{$\sim$}}}}
\def \gta {\mathrel{\vcenter
     {\hbox{$>$}\nointerlineskip\hbox{$\sim$}}}}

\def\Red{}
\def\Black{}
\def\Blue{}

\newcommand{\lascia}[1]{}
\def\puttag(#1,#2)#3{\put(#1,#2){\makebox(0,0){\rm\Blue #3\Black}}}

\def\circa#1{\,\raise.3ex\hbox{$#1$\kern-.75em\lower1ex\hbox{$\sim$}}\,}

\newcommand{\psfigure}[1]{\ifnum\Mac=1 \special{picture #1} \else
\includegraphics{#1} \fi}
\def\putps(#1,#2)(#3,#4)#5#6{\ifnum\Mac=1 \put(#1,#2){\special{picture #5}}
\else  \put(#3,#4){\includegraphics{#6}} \fi}

\def\One{\hbox{1\kern-.24em I}}

\newcommand\Ord{{\cal O}}

\newcommand{\eq}[1]{~{\rm (\ref{eq:#1})}}

\makeatletter
\def\art{\@ifnextchar[{\eart}{\oart}}
\def\eart[#1]#2#3#4#5#6{{\rm #2}, {\e, #3 \bf #4} {\rm (#6) #5} ({\em #1})}
\def\hepart[#1]#2{{\rm #2, \em#1}}
\newcommand{\oart}[5]{{\rm #1}, {\em #2 \bf #3} {\rm (#5) #4}}
\newcounter{alphaequation}[equation]
\def\thealphaequation{\theequation\hbox to
0.6em{\hfil\alph{alphaequation}\hfil}}
\def\eqnsystem#1{
\def\@eqnnum{{\rm (\thealphaequation)}}
\def\@@eqncr{\let\@tempa\relax \ifcase\@eqcnt \def\@tempa{& & &} \or
  \def\@tempa{& &}\or \def\@tempa{&}\fi\@tempa
  \if@eqnsw\@eqnnum\refstepcounter{alphaequation}\fi
\global\@eqnswtrue\global\@eqcnt=0\cr}
\refstepcounter{equation} \let\@currentlabel\theequation \def\@tempb{#1}
\ifx\@tempb\empty\else\label{#1}\fi
\refstepcounter{alphaequation}
\let\@currentlabel\thealphaequation
\global\@eqnswtrue\global\@eqcnt=0 \tabskip\@centering\let\\=\@eqncr
$$\halign to \displaywidth\bgroup \@eqnsel\hskip\@centering
$\displaystyle\tabskip\z@{##}$&\global\@eqcnt\@ne
\hskip2\arraycolsep\hfil${##}$\hfil& \global\@eqcnt\tw@\hskip2\arraycolsep
$\displaystyle\tabskip\z@{##}$\hfil
\tabskip\@centering&\llap{##}\tabskip\z@\cr}

\def\endeqnsystem{\@@eqncr\egroup$$\global\@ignoretrue} \makeatother

\begin{document}\small

\twocolumn[
\begin{quote}
{\em August 1998}\hfill {\bf SNS-PH/98-13}\\
hep-ph/9808263 \hfill{\bf IFUP-TH/98-26}
\end{quote}
\medskip
\begin{center}
{\Large\bf\Red The Region of Validity of Homogeneous Nucleation Theory}\\
\bigskip\bigskip\Black\normalsize
{\bf  A. Strumia$^{\rm (a)}$,
N. Tetradis$^{\rm (b)}$ {\rm and} C. Wetterich$^{\rm (c)}$}\\
\bigskip
\normalsize{\em
(a) Dipartimento di Fisica, Universit\`a di Pisa and
INFN, sezione di Pisa, I-56127 Pisa, Italy\\
(b) Scuola Normale Superiore, Piazza dei Cavalieri 7, I-56126 Pisa, Italy\\
(c) Institut f\"ur Theoretische Physik, Universit\"at Heidelberg,
Philosophenweg 16, D-69120 Heidelberg, Germany}\\
\bigskip
\Blue{\large\bf Abstract}
\end{center}
\begin{quote}\normalsize\indent
We examine the region of validity of Langer's picture of
homogeneous nucleation.
Our approach is based on a coarse-grained free energy
that incorporates the effect of fluctuations with momenta
above a scale $k$.
The nucleation rate $I=A_k \exp(-S_k)$
is exponentially suppressed by the action $S_k$ of the
saddle-point configuration that dominates tunnelling.
The factor $A_k$ includes a fluctuation determinant around this saddle point.
Both $S_k$ and $A_k$ depend on the choice of $k$, but, for
$1/k$ close to the characteristic length scale of the saddle point,
this dependence cancels in the expression for the nucleation rate.
For very weak first-order phase transitions or in the vicinity of the
spinodal decomposition line,
the pre-exponential factor $A_k$ compensates
the exponential suppression $\exp(-S_k)$.
In these regions the standard nucleation picture breaks down.
We give an approximate expression for $A_k$ in terms of the saddle-point
profile, which can be used for quantitative estimates and
practical tests of the validity of homogeneous nucleation theory.
\end{quote}\Black
\vspace{1cm}~]

\paragraph{1. Introduction:}
The theory of first-order phase transitions
is a subject of much interest to
statistical and particle physicists (for a review see ref.
\cite{review} and references therein).
Our present understanding of these phenomena is based on
the work of Langer on homogeneous nucleation theory~\cite{langer}.
His formalism has been applied to relativistic field theory by
Coleman  and Callan~\cite{coleman} and extended
to thermal equilibrium by
Affleck and Linde~\cite{affleck}.
The basic quantity in this approach is the nucleation rate
$I=A \exp(-S)$, which
gives the probability per unit time and volume to nucleate a
region of the stable phase (the true vacuum) within the metastable
phase (the false vacuum).
For a strong enough first-order transition
the rate is exponentially suppressed by the free energy of
the critical bubble, which is a static configuration
(usually assumed to be spherically symmetric) within the metastable phase,
whose interior consists of the stable phase. Bubbles larger
than the critical one expand rapidly, thus converting the
metastable phase into the stable one.
Deformations of the critical  bubble in the thermal bath
generate  a fluctuation determinant around the critical-bubble
configuration which is contained in $A$.
Another dynamical prefactor determines the fast growth rate
of bubbles larger than the critical one~\cite{langer,kapusta}.
We concentrate here on the calculation of the
static prefactor.

Up to the dynamical prefactor, the bubble-nucleation rate
is $I\sim \exp \left\{-  \left( \Gamma
\left[ \phi_b(r) \right]-\Gamma\left[ \phi_f \right] \right)\right\}$,
with $\Gamma$ the free energy, evaluated either at the false vacuum
where $\Gamma$ has a local minimum for a constant field $\phi_f$,
or for the inhomogeneous field configuration $\phi_b(r)$
which interpolates between the two
vacua and is a saddle point of $\Gamma$. This configuration is
usually identified with the critical bubble.
The problem of computing $\Gamma[\phi_b]$ may be
divided into three steps:
In the first step, one only includes fluctuations with momenta
larger than a scale $k$ which is of the order of the typical gradients
of $\phi_b(r)$. For this step one can consider approximately
constant fields $\phi$ and use a derivative expansion for the resulting
coarse-grained free energy
$\Gamma_k[\phi]$. The second step searches for the
configuration $\phi_b(r)$ which is a saddle point of $\Gamma_k$.
Finally, the remaining fluctuations with momenta smaller
than $k$ are evaluated in a saddle-point approximation
around $\phi_b(r)$. This procedure
of coarse graining systematically avoids problems with
double-counting the effect of fluctuations,
the convexity of the effective potential that determines
part of $\Gamma[\phi_b]$,
or the ultraviolet regularization of the fluctuation determinants in $A$
\cite{rates}.
As a test of the validity of the approach, the result for the rate
$I$ must be independent of the coarse-graining scale $k$, as
the latter should be considered only as a technical device.
Langer's approach corresponds to a one-loop approximation
around the dominant saddle point
for fluctuations with momenta smaller than $k$, whereas the
coarse-grained free energy $\Gamma_k$ is often chosen
phenomenologically.

\paragraph{2. The method:}
We employ the formalism of the effective average action
$\Gamma_k$~\cite{exact}, which can be identified with the
free energy at a given coarse-graining scale
$k$. In the limit $k \to 0$, $\Gamma_k$ becomes equal to the
effective action. We consider a
statistical system with one space-dependent degree of freedom described
by a real scalar field $\phi(x)$.
For example, $\phi(x)$ may correspond to
the density for the gas/liquid transition,
or to a difference in concentrations for chemical phase transitions,
or to magnetization for the ferromagnetic transition.
Our discussion also applies to a quantum field theory in
thermal equilibrium for scales $k$ below the temperature $T$.
Then an effective three-dimensional description~\cite{trans}
applies and we assume that
$\Gamma_{k_0}$ has been computed (for example perturbatively) for $k_0=T$.

We approximate $\Gamma_k$ by a standard kinetic term and
a general potential $U_k$.
This is expected to be a good approximation, because the size of
the higher-derivative terms
is related to the anomalous dimension of the field,
which is small for this model ($\eta \simeq 0.035$).
At a short-distance scale $k_0^{-1}=T^{-1}$, the long-range
collective fluctuations
are not yet important and we assume a potential
\be
U_{k_0} (\phi) =
\frac{1}{2}m^2_{k_0} \phi^2
+\frac{1}{6} \gamma_{k_0} \phi^3
+\frac{1}{8} \lx_{k_0} \phi^4.
\label{eq:two20} \ee
The parameters $m_{k_0}^2, \gamma_{k_0}$ and $\lambda_{k_0}$
depend on $T$. This potential has the typical form relevant for
first-order phase transitions in four-dimensional
field theories at high temperature.
Through a shift $\phi\to\phi+c$ the cubic term can be eliminated
in favour of a term linear in $\phi$ \cite{B,rates}.
The potential (\ref{eq:two20}) therefore
describes statistical systems of the Ising universality class
in the presence of an external magnetic field.
For a Hamiltonian
\be
H=\int d^3x\,
\left\{
\frac{\hat{\lx}}{8}\left(\chi^2-1\right)^2-B\chi+\frac{\zeta}{2}
\,\partial_i\chi\,\partial^i\chi
\right\},
\label{hamil} \ee
the parameters are
$m^2_{k_0}=\hat{\lx}(3y^2-1)/2\zeta$, $\gamma_{k_0}=3\hat{\lx}T^{1/2}y/
\zeta^{3/2}$, $\lx_{k_0}=\hat{\lx}T/\zeta^2$, with $y$ given by
$y(y^2-1)=2B/\hat{\lx}$.
For real magnets $k_0$ must be taken somewhat
below the inverse lattice distance,
so that effective rotation and translation symmetries apply.
Correspondingly, $\chi$ and $H$ are the effective normalized spin field and
the effective Hamiltonian at this scale.
Our choice of potential encompasses a large class of field-theoretical
and statistical systems.
In a different context our results can also
be applied to the problem of quantum tunnelling in a (2+1)-dimensional
theory at zero temperature. In this case $k_0,m^2,\gamma$ and
$\lambda$ bare no relation to temperature.

We compute the form of the potential $U_k$ at scales $k\leq k_0$ by
integrating an evolution equation~\cite{trans}.
The latter is derived from an exact flow
equation for $\Gamma_k$~\cite{exact}, typical of the Wilson approach to the
renormalization group~\cite{wilson}. The form of $U_k$ changes as
the effect of fluctuations with momenta above the decreasing scale
$k$ is incorporated in
the effective couplings of the theory.
We consider an arbitrary form of $U_k$ which, in general,
is not convex for non-zero $k$.
$U_k$ approaches the convex effective potential only in the limit $k\to 0$.
In the region relevant for a first-order phase
transition, $U_k$ has two distinct
local minima.
The nucleation rate should be computed for $k$ larger than
or around the scale $k_f$ at which  $U_k$ starts  receiving important
contributions from field configurations that interpolate between
the two minima. This happens when the negative curvature at the top
of the barrier becomes approximately equal to $-k^2$~\cite{convex}.
Another consistency check for the above choice of $k$
is provided by the fact that
for $k > k_f$ the typical length scale of a thick-wall critical
bubble is $\gta  1/k$.

We use here a mass-like infrared cutoff $k$ for the fluctuations
that are incorporated in $\Gamma_k$ and neglect
the anomalous dimension.
The evolution equation for the potential is~\cite{exact,trans,rates}
$$\frac{\partial }{\partial k^2}\left[U_k(\phi)-U_k(0)\right]= \qquad\qquad\qquad$$
\be\qquad\qquad\label{eq:2}
=- \frac{1}{8 \pi}\left[\sqrt{k^2+U_k''(\phi)}
-\sqrt{k^2+U_k''(0)}\right]. \label{twofour} \ee
It is instructive to compare with the first
step of an iterative solution of the general flow equation~\cite{A}
\begin{equation}\label{eq:4}
U_k^{(1)}(\phi)-U_k^{(1)}(0)=U_{k_0}(\phi)-U_{k_0}(0)+\qquad\qquad
\end{equation}
$$+
\frac{1}{2}\ln\left[\frac{\det[-\partial^2+k^2+U_k''(\phi)]}{
\det[-\partial^2+k^2_0+U_k''(\phi)]}
\frac{\det[-\partial^2+k^2_0+U_k''(0)]}{
\det[-\partial^2+k^2+U_k''(0)]}\right].$$
For $k\to 0$ this is a regularized one-loop
approximation to the effective
potential. Due to the ratio of determinants, only
momentum modes with $k^2<q^2<k_0^2$ are effectively included
in the momentum integrals in\eq{4}.
(Eq.~(\ref{eq:2}) can be derived formally from eq.~(\ref{eq:4})
by performing the
momentum integration, i.e.\ $\ln\det F(q^2)=\int \ln F(q^2)~d^3q/(2\pi)^3$,
and taking the derivative $\partial/\partial k^2$ assuming that it
does not act on $U_k$).

\medskip

The nucleation rate  per unit volume $I$
(probability of nucleation of a critical bubble
per unit time and volume)
is $I=A_k\exp(-S_k)$, where
$S_k=\int d^3r[-\frac{1}{2}\phi_b(r)\Delta \phi_b(r)+U_k(\phi_b(r))]$
is evaluated for the bubble solution $\phi_b(r)$
which is a saddle point of $S_k$ and interpolates
between the true and the false vacuum. The pre-exponential factor is
$$A_k = \frac{E_0}{2\pi}\left( \frac{S_k}{2\pi}\right)^{3/2}
\bigg[
\frac{\det'\left[-\partial^2+U''_k(\phi_b(r))\right]}
{\det \left[ -\partial^2+k^2+U''_k(\phi_b(r))\right]}\times$$
\begin{equation} \times
\frac{\det\left[-\partial^2+k^2+U''_k(0)\right]}
{\det\left[-\partial^2+U''_k(0)\right]}
\bigg]^{-1/2}.
\label{eq:rrate}
\end{equation}
Eq.~(\ref{eq:rrate})
is the standard expression for
the nucleation rate~\cite{langer,affleck} with fluctuation determinants
replaced by ratios of determinants, in complete analogy to eq.~(\ref{eq:4}).
This ensures that only fluctuations with momenta $q^2 \lta k^2$ are included
in $A_k$. One observes that, up to the difference between the saddle point
$\phi_b(r)$ and the constant field $\phi$, the explicit $k$-dependence of
$-\ln I$ cancels between the one-loop contribution to
$U_k$ (see eq.\eq{4}) and $-\ln A_k$.
The prime in the fluctuation determinant around
the saddle point denotes that the three zero eigenvalues are not included.
Their contribution generates the factor
$(S_k/2\pi)^{3/2}$ and the volume factor
that is absorbed in the definition of $I$.
The quantity $E_0$ is the square root of
the absolute value of the unique negative eigenvalue.
This last contribution appears only for the high-temperature
field theory~\cite{affleck}. It is absent in the expression
for the quantum-tunnelling rate
in the zero-temperature (2+1)-dimensional theory.

The coarse-grained potential $U_k$ is
determined through the numerical integration
of eq.~(\ref{twofour}) between the scales $k_0$ and $k$,
using, for example, algorithms from ref.~\cite{num}.
The initial condition for the integration
is given by eq.~(\ref{eq:two20}).
The computation of $S_k$ and $A_k$ is presented in detail in
ref.~\cite{rates}, where techniques from
refs.~\cite{cott} are adopted.
In contrast to refs.~\cite{cott}, our calculation is ultraviolet finite and
no additional regularization is needed.

A possible $k$-dependence of the final result for the nucleation
rate may result from three sources: The first is an insufficiency
of the one-loop approximation for $\ln A_k$ which may
not match with the more precise non-perturbative
determination of $S_k$. This error grows with increasing $k$.
The second error comes from the replacement of $\phi_b(r)$
by a slowly varying field in the computation of $U_k$ and $S_k$.
It increases with decreasing $k$ and may become substantial for
$k<k_f$. Finally, the critical bubble $\phi_b(r)$ is determined
by an extremization procedure which does not take into account the
contributions to the free energy from fluctuations with momenta
smaller than $k$.
It is certainly a necessary requirement
for the validity of Langer's nucleation theory that the $k$-dependence
of the nucleation rate comes out small in some appropriate range of
$k$~\cite{bubble1}
(typically near $k_f$). Furthermore, nucleation theory will
break down if the rate ceases to be exponentially suppressed.
This typically happens near
the spinodal line (along which the false vacuum becomes unstable),
and, in particular, in the vicinity of
the endpoint of the first-order
critical line --- in our case at a second-order phase transition.
For the potential of eq.~(\ref{eq:two20}),
the two first-order critical lines obey
$\gamma_{k_0}^2=9\lambda_{k_0} m_{k_0}^2$ and
$\gamma_{k_0}=0$, with endpoints at $m_{k_0}^2=-2\mu_{\rm cr}^2$,
$\gamma_{k_0}^2=-18\lambda_{k_0}\mu_{\rm cr}^2$ and
$m_{k_0}^2=\mu_{\rm cr}^2$, $\gamma_{k_0}=0$~\cite{B}.
Here $\mu_{\rm cr}^2$ is the critical mass term of the Ising model
($\mu_{\rm cr}^2/k_0^2 \approx -0.0115$ for $\lambda_{k_0}/k_0=0.1$).
We point out that, for fixed $m_{k_0}^2$ and $\lambda_{k_0}$, opposite
values of $\gamma_{k_0}$ result in potentials related through
$\phi \leftrightarrow -\phi$.
Also a model with $m_{k_0}^2<0$ can be mapped onto the equivalent
model with $m^{\prime2}_{k_0}>0$ by the shift $\phi\to\phi+c$,
$\lambda_{k_0}c^2+\gamma_{k_0}c=-2m_{k_0}^2$, where
$m^{\prime 2}_{k_0}=-2m_{k_0}^2-\frac{1}{2}\gamma_{k_0}c$,
$\gamma'_{k_0}=\gamma_{k_0}+3\lambda_{k_0}c$.

\begin{figure*}[t]
\begin{center}
\begin{picture}(18,12)
\putps(-0.7,-0.4)(-2.2,-0.5){f1}{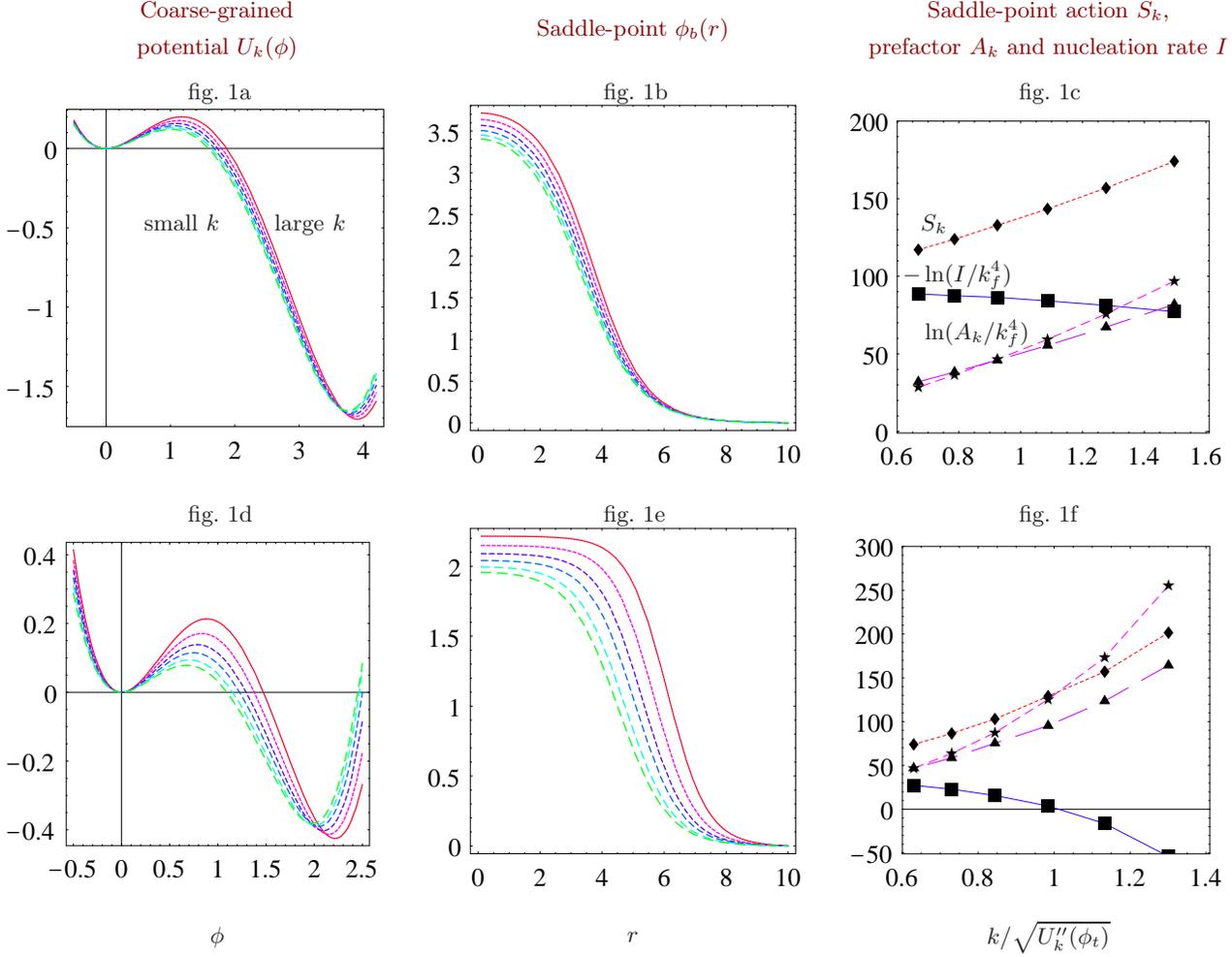}\Red
\put(2.6,12){\makebox(0,0){Coarse-grained}}
\put(2.6,11.5){\makebox(0,0){potential $U_k(\phi)$}}
\put(8.3,11.75){\makebox(0,0){Saddle-point $\phi_b(r)$}}
\put(14,12){\makebox(0,0){Saddle-point action $S_k$,}}
\put(14,11.5){\makebox(0,0){prefactor $A_k$ and nucleation rate $I$}}
\Black
\put(2.6,-0.7){\makebox(0,0){$\phi$}}
\put(8.3,-0.7){\makebox(0,0){$r$}}
\put(14,-0.7){\makebox(0,0){$k/\sqrt{U_k''(\phi_t)}$}}
\put(2.2,10.8){fig.~1a}\put(7.9,10.8){fig.~1b}\put(13.6,10.8){fig.~1c}
\put(2.2,5){fig.~1d}\put(7.9,5){fig.~1e}\put(13.6,5){fig.~1f}
\put(1.6,9){\small small $k\qquad$ large $k$}
\put(12,9){$\phantom{-}S_k$}
\put(12,8.3){$-\ln(I/k_f^4)$}
\put(12,7.5){$\phantom{-}\ln (A_k/k_f^4)$}
\end{picture}
\vspace{7mm}
\caption[SP]{\em Dependence of effective potential, critical bubble and
nucleation rate on the coarse graining scale $k$.
The parameters are $\lambda_{k_0}=0.1\cdot k_0$,
$m^2_{k_0}=-0.0433\cdot k_0^2$, $\gamma_{k_0}=-0.0634~k_0^{3/2}$
(figs. 1a--1c)
and
$m^2_{k_0}=-0.013\cdot k_0^2$, $\gamma_{k_0}=-1.61 \cdot 10^{-3}~k_0^{3/2}$
(figs. 1d--1f).
All dimensionful quantities are given in units of $k_f$,
equal to $0.223\cdot k_0$ in the first series
and to  $0.0421\cdot k_0$ in the second series.
\label{fig:1}}
\end{center}\end{figure*}

\paragraph{3. Sample computations:}
We present here a
computation of the spontaneous nucleation rate and establish
the region of validity of Langer's theory.
Fig.~1 exhibits the results of our calculation
for the potential~(\ref{eq:two20})
with
$m^2_{k_0}=-0.0433~k_0^2$,
$\gamma_{k_0}=-0.0634~k_0^{3/2}$,
$\lx_{k_0}=0.1~k_0$. We first show in
fig.~1a  the evolution of the potential $U_k(\phi)$ as the scale
$k$ is lowered.
(We always shift the metastable vacuum to $\phi=0$.)~
The solid line corresponds to $k/k_0=0.513$ while the
line with longest dashes (that has the smallest barrier height)
corresponds to $k_f/k_0=0.223$. At the scale $k_f$ the negative
curvature at the top of the barrier is slightly larger than
$-k_f^2$ and we stop the evolution.
The potential and the field have been
normalized with respect to $k_f$.
As $k$ is lowered from $k_0$ to $k_f$, the absolute minimum of the potential
settles at a non-zero value of $\phi$, while a significant barrier
separates it from the metastable minimum at $\phi=0$.
The profile of the critical bubble $\phi_b(r)$
is plotted in fig.~1b in units of $k_f$
for the same sequence of scales.  For $k\simeq k_f$ the characteristic
length scale of the bubble profile and $1/k$ are comparable. This is expected,
because the form of the profile is determined by the barrier of the potential,
whose curvature is $\simeq -k^2$ at this point.
This is an indication that we should not proceed to coarse-graining
scales below $k_f$.
We observe a significant
variation of the value of the field $\phi$ in the interior of the bubble
for different $k$.

Our results for the nucleation rate are presented in fig.~1c.
The horizontal axis corresponds to $k/\sqrt{U''_k(\phi_t})$,
i.e. the ratio of the scale $k$
to the square root of the positive curvature (equal to the
mass of the field) at the
absolute minimum of the potential located at $\phi_t$.
Typically, when $k$ crosses below this mass,
the massive fluctuations of the field
start decoupling. The evolution of the convex parts of
the  potential slows down and eventually stops.
The dark diamonds give the values of the action $S_k$ (free energy
rescaled by the temperature)
of the critical bubble at the scale $k$. We observe a strong
$k$ dependence of this quantity, which is expected from
the behaviour in figs.~1a, 1b.
The stars in fig.~1c indicate the values of
$\ln ( A_k/k^4_f )$.
Again a substantial decrease with decreasing $k$ is observed. This is expected,
because $k$ acts as the effective ultraviolet cutoff in the calculation
of the fluctuation determinants in $A_k$.
The dark squares give our results for
$-\ln(I/k^4_f )
= S_k-\ln ( A_k/k^4_f )$. It is remarkable that the
$k$ dependence of this quantity almost disappears for $k/\sqrt{U''_k(\phi_t})
\circa{<} 1$.
The small residual dependence on $k$ can be used to estimate the
contribution of the next order in the expansion around the saddle point.
It is reassuring that this contribution is expected to be
smaller than $\ln ( A_k/k^4_f )$.

This behaviour confirms our expectation that the
nucleation rate should be independent of the scale $k$ that
we introduced as a calculational tool. It also demonstrates that
all the configurations plotted in fig.~1b give equivalent
descriptions of the system, at least for the lower values of $k$.
This indicates that the critical bubble should not be associated only
with the saddle point of the semiclassical approximation, whose
action is scale dependent. It is the combination of
the saddle point and its possible deformations
in the thermal bath that has physical meaning.
We point out that a reliable calculation of the nucleation rate
is only possible if the ultraviolet cutoff in the fluctuation determinant
matches properly with the infrared scale in the coarse-graining procedure.
This problem was not quantitatively accessible before the present work.

For smaller values of $|m^2_{k_0}|$ the dependence of the nucleation
rate on $k$ becomes more pronounced. We demonstrate this in the
second series of figs.~1d--1f where
$(-m^2_{k_0})^{1/2}/\lambda_{k_0}$ $=1.13$
(instead of $2.08$ for figs.~1a--1c).
The value of $\lambda_{k_0}$ is the same as before, whereas
$\gamma_{k_0}=-1.61\cdot 10^{-3}k_0^{3/2}$ and $k_f/k_0=0.0421$.
The reason for this behaviour is the larger value of the dimensionless
renormalized quartic coupling~\cite{bubble1} for the second parameter set.
Higher loop contributions to $A_k$ become more important.

\smallskip

It is apparent from figs.~1c and 1f
that the leading contribution to the
pre-exponential factor increases the total nucleation rate.
This behaviour, related to the fluctuations of the field whose expectation
value serves as the order parameter, is observed
in multi-field models as well.
The reason can be traced to the form of the differential
operators in the prefactor of eq.~(\ref{eq:rrate}).
This prefactor involves the ratio
$\det'\left[-\partial^2+U''_k(\phi_b(r))\right]/
\det\left[-\partial^2+U''_k(0)\right]$ before regularization.
The function $U''_k(\phi_b(r))$ always has a minimum away from
$r=0$ where it takes negative values (corresponding to
the negative curvature at the top of the barrier), while
$U''_k(0)$ is always positive. As a result the lowest
eigenvalues of the operator $\det'\left[-\partial^2+U''_k(\phi_b(r))\right]$
are smaller
than those of $\det\left[-\partial^2+U''_k(0)\right]$. The elimination of
the very large eigenvalues from the determinants
through regularization does not affect this
fact and the prefactor is always larger than 1. Moreover,
for weak first-order phase transitions it becomes exponentially large
because of the proliferation of low eigenvalues of the first operator.
In physical terms, this implies the
existence of a large class of field configurations of free energy comparable
to that of the saddle-point. Despite the fact that they are not
saddle points of the free energy
(they are rather deformations of a saddle point)
and are, therefore, unstable, they result in a dramatic increase of
the nucleation rate.
This picture is very similar to that of
``subcritical bubbles'' of ref.~\cite{gleiser}.

\paragraph{4. Approximate expression for the prefactor:}
In figs.~1c and 1f we also display the values of
$\ln ( A_k/k^4_f )$ (dark triangles) predicted by the approximate expression
\begin{equation}
\ln \frac{A_k}{k^4_f}
\approx \frac{\pi k}{2}
\left[
- \int_0^\infty \!\!\! r^3 \left[
U''_k\left( \phi_b(r) \right)
-U''_k\left( 0 \right)
\right] dr
\right]^{1/2}\equiv D\pi.
\label{eq:appr}
\end{equation}
This expression is obtained as follows:
The prefactor, given by a combination of determinants involving
the Laplacian operator,
can be written as a product of contributions with fixed angular
momentum number $\ell$:
$A_k=\prod_\ell c_\ell$.
For large $\ell$, the factors $c_\ell$ can be evaluated analytically
as $c_\ell = 1+D^2/\ell^2+\Ord(\ell^{-4})$ \cite{rates}.
We have checked that the first two terms of this expansion give a
good approximation to $c_\ell$ even for
$\ell\circa{<}D$,
except for $\ell = \Ord(1)$. Unless the prefactor is of order 1, we obtain
$A_k\approx\prod_\ell c_\ell\approx e^{D\pi}$.

\begin{figure}[t]
\begin{center}\hspace{-5mm}
\begin{picture}(9,5)
\putps(-0.5,0)(-0.5,0){fTh}{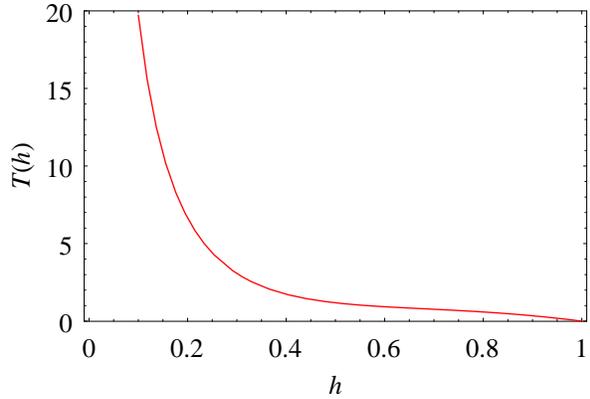}
\end{picture}
\caption[SP]{\em The parameter $T(h)$, defined in eq.~$(\ref{fin})$,
as a function of $h.$
\label{fig:SAh}}
\end{center}\end{figure}

Eq.\eq{appr} permits
a quick evaluation of the prefactor in terms of the bubble profile.
It is useful to obtain some intuition on the behaviour of the
nucleation rate by using this expression.
We assume that the potential has an approximate form similar
to eq.~(\ref{eq:two20}) even near $k_f$, i.e.
\be
U_{k_f} (\phi) \approx
\frac{1}{2}m^2_{k_f} \phi^2
+\frac{1}{6} \gamma_{k_f} \phi^3
+\frac{1}{8} \lx_{k_f} \phi^4,
\label{eq:two21} \ee
with $m^2_{k_f} > 0$.
For systems
not very close to the endpoint of the first-order critical line,
this assumption is supported by
the numerical data, as can be verified from fig.~1.
The scale $k_f$ is determined by the relation
\be
k^2_f \approx \max \left| U''_{k_f}(\phi)
\right|=\frac{\gamma_{k_f}^2}{6\lx_{k_f}}-m_{k_f}^2.
\label{abc} \ee
Through the rescalings
$r=\rt/m_{k_f}$, $\phi=2\phit\, m_{k_f}^2/\gamma_{k_f}$, the potential
can be written as
$$\Ut(\phit)=\frac{1}{2}\phit^2-\frac{1}{3}\phit^3+\frac{h}{18}\phit^4,$$
with $h=9\lx_{k_f} m_{k_f}^2/\gamma_{k_f}^2$.
For $h \approx 1$
the two minima of the potential have approximately
equal depth. The action of the saddle point can be expressed
as
\be
S_{k_f} =
\frac{4}{9} \frac{m_{k_f}}{\lx_{k_f}}\, h \St(h),
\label{action2} \ee
where $\St(h)$ must be determined numerically through $\Ut(\phit)$.
Similarly, the pre-exponential factor can be estimated through
eq.~(\ref{eq:appr}) as
\beq
\ln \frac{ A_{k_f}}{k^4_f}  &\approx&
\frac{\pi}{2}
\sqrt{ \frac{3}{2h}-1}\,\, \At(h),
\nonumber \\
\At^2(h) &=&
\int_0^\infty \Ut'' \left(\phit_b(\rt)\right) \,\rt^3\, d\rt,
\label{prefactor2} \eeq
with $\At(h)$ computed numerically.
Finally
\be
R=\frac{\ln(A_{k_f}/k_f^4)}{S_{k_f}} \approx
\frac{9\pi}{8h}\sqrt{\frac{3}{2h}-1}\,
\frac{\At(h)}{\St(h)}
\,\frac{\lx_{k_f}}{m_{k_f}}
=T(h) \,\frac{\lx_{k_f}}{m_{k_f}}.
\label{fin} \ee
In fig.~2 we plot $T(h)$ as a function
of $h$ in the interval (0,\,1).
It diverges for $h\to 0$.
For $h \to 1$, our estimate of the prefactor
predicts $T(h)\to 0$. The reason is that, for our approximate
potential of eq.~(\ref{eq:two21}), the field masses
at the two minima are equal in this limit. As a result, the
integrant in eq.~(\ref{eq:appr}) vanishes, apart from at the surface
of the bubble. The small surface contribution is negligible for
$h\to 1$, because the
critical bubbles are very large in this limit.
This behaviour is not expected to
persist for more complicated potentials. Instead, we expect a
constant value of $T(h)$ for $h\to 1$.
However, the approximate expression~(\ref{eq:appr}) has not
been tested for very large critical bubbles. The divergence of the
saddle-point action in this limit results in low accuracy
for our numerical analysis. Typically, our results are reliable
for $\St(h)$ less than a few thousand.
Also, eq.~(\ref{eq:appr})
relies on a large-$\ell$ approximation.
For increasing $D$, this
breaks down below an increasing value $\ell_{\rm as}$ and, therefore,
eq.~(\ref{eq:appr}) is not guaranteed to be valid.
We have checked that
both our numerical and approximate results are reliable for
$h \lta 0.9$.

The estimate of eq.~(\ref{fin}) suggests two cases in which
the expansion around the saddle point is expected to break down:
\begin{itemize}

\item[a)]For fixed $\lx_{k_f}/m_{k_f}$, the ratio $R$ becomes larger than 1 for
$h\to 0$. In this limit the barrier becomes negligible and the
system is close to the spinodal line. 

\item[b)] For fixed $h$, again $R$ can be large for sufficiently
large $\lx_{k_f}/m_{k_f}$.

\end{itemize}
Case (b) is possible even for $h$ close
to 1, so that the system is far from the spinodal line. This
case corresponds to weak first-order phase transitions, as
can be verified by observing that
the saddle-point action of eq.~(\ref{action2}),
the location of the true vacuum
\be
\frac{\phi_t}{\sqrt{m_{k_f}}}=
\frac{2}{3}\sqrt{h}\,\,
\phit_t(h)\,\sqrt{\frac{m_{k_f}}{\lx_{k_f}}},
\label{phit} \ee
and the
difference in
free-energy density between the minima
\be
\frac{\Delta U}{m_{k_f}^3}=
\frac{4}{9}\,{h}\,\,
\Delta\Ut(h)\,\frac{m_{k_f}}{\lx_{k_f}}
\label{deltaU} \ee
go to zero in the limit $m_{k_f}/\lx_{k_f}\to 0$ for fixed $h$. This
is in agreement with the discussion of fig.~1 in the previous section.\\
The breakdown of homogeneous nucleation theory in both the above
cases is confirmed through the numerical computation of the nucleation rates.

\medskip

\paragraph{5. The region of validity of homogeneous nucleation theory:}
In fig.~3 we show contour plots for $I/k_f^4$ and for
$R=\ln(A_{k_f}/k^4_f)/S_{k_f}$ in the $(m_{k_0}^2,\gamma_{k_0})$
plane for fixed $\lambda_{k_0}/k_0=0.1$.
One can see the decrease of the rate as
the first-order critical line $\gamma_{k_0}=0$ is approached.
The spinodal line (end of the shaded region),
on which the metastable minimum of $U_k$ becomes unstable, is also
shown.
The nucleation rate becomes large before the
spinodal line is reached.
For $-\ln (I/k_f^4)$ of order 1,
the exponential suppression of the nucleation
rate disappears. Langer's approach can no longer be applied
and an alternative picture for the dynamical transition must be
developed~\cite{boy}.
In the region between the contour $I/k_f^4=e^{-3}$ and the
spinodal line, one expects a smooth transition from nucleation to
spinodal decomposition.
The spinodal and critical lines meet at the endpoint in the
lower right corner, which corresponds to a second order phase transition.
The figure exhibits an increasing rate as
the endpoint is approached at a fixed distance from the critical line.

The ratio $R$ is a measure
of the validity of the semiclassical approximation. For $R\approx 1$
the fluctuation determinant is as important as the ``classical''
exponential factor $e^{-S_k}$. There is no reason to assume that
higher loop contributions from the fluctuations around the critical
bubble can be neglected anymore. Near the endpoint in the lower right corner,
Langer's semiclassical picture
breaks down, despite the presence of a discontinuity in the order
parameter. Requiring $I/k_f^4\circa{<}e^{-3}$, $R\circa{<}1$,
gives a limit of validity for Langer's theory.
For a fixed value of the nucleation rate (solid lines in fig.~3),
the ratio $R$ grows as the endpoint in the lower right corner is
approached. This indicates that Langer's theory is not applicable
for weak first-order phase transitions, even if the predicted rate
is exponentially suppressed. The concept of nucleation of a region of
the stable phase within the metastable phase may still be relevant. However,
a quantitative estimate of the nucleation rate requires taking into account
fluctuations of the system that are not described properly by the
semiclassical approximation \cite{gleiser}.

The parameter region discussed
here may be somewhat unusual since the critical line
of the phase
transition is approached  by varying $\gamma_{k_0}$ from negative or positive
values
towards zero. We have chosen it only for making the graph more
transparent.
However,
the results of fig.~3 can be mapped by a shift $\phi\to\phi+c$ to another
region with $m_{k_0}^2>0$, for which the
first-order phase transition can be approached by varying $m_{k_0}^2$
at fixed $\gamma_{k_0}$ (see end of section 2).
As opposite values of $\gamma_{k_0}$ result in potentials
related by
$\phi \leftrightarrow -\phi$, we can always choose
$\gamma_{k_0}<0$. Then the phase transition proceeds
from a metastable minimum at the origin to a stable minimum
along the positive $\phi$-axis (as in fig.~1a). Potentials with
$m_{k_0}^2>0$, $\gamma_{k_0}<0$
are relevant for cosmological phase transitions, such as the
electroweak.

\begin{figure}[t]\begin{center}\begin{picture}(7,7)
\putps(-0.5,-1)(-0.5,-1){f2}{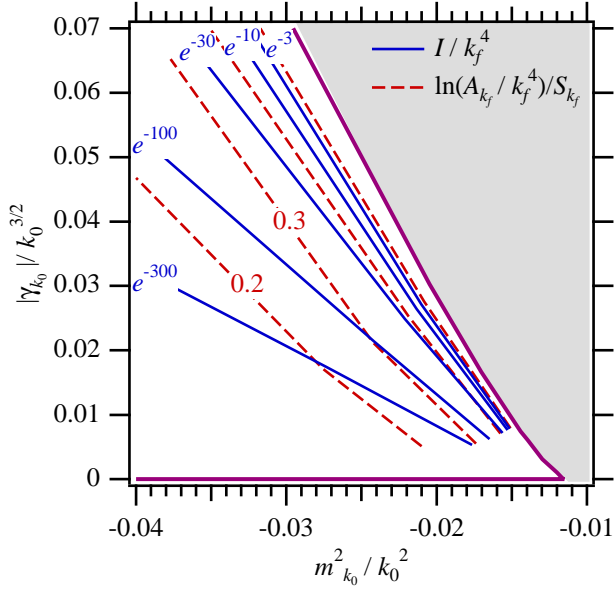}
\end{picture}\vspace{7mm}
\caption[SP]{\em Contour plots of $I/k_f^4$
and of $R=\ln(A_{k_f}/k_f^4)/S_{k_f}$  in the plane
$(m_{k_0}^2,\gamma_{k_0})$, for $\lx_{k_0}/k_0=0.1$.
Regions to the right of the spinodal line
(only one minimum) are shaded.
The dashed lines correspond to $R=\{0.2,0.3,0.5,1\}$ and the solid lines
to $I/k_f^4$.\label{fig:2}}
\end{center}\end{figure}

\begin{figure}[t]\begin{center}\begin{picture}(7,7)
\putps(-0.5,-1)(-0.5,-1){f2V}{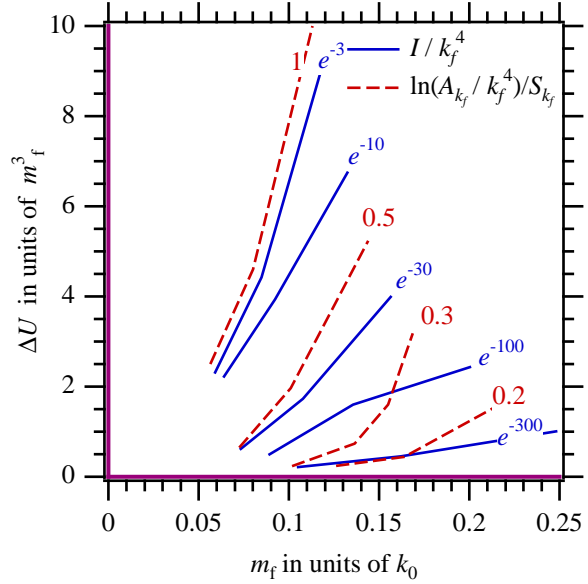}
\end{picture}\vspace{7mm}
\caption[SP]{\em The contour plots of fig.~3
in the plane $(m_{f},\Delta U)$, where $m_f$ is the mass at the false vacuum and $\Delta U$ is
the difference in free-energy density between the two vacua.
The spinodal line corresponds to the vertical axis.
\label{fig:3}}
\end{center}\end{figure}

\smallskip

In fig.~4 we depict the region of validity of homogeneous nucleation
theory in terms of parameters of the low-energy theory at the scale
$k_f$. The contours
correspond to the same quantities as in fig.~3, which are now
plotted as a function of the renormalized mass at the false
vacuum $m_f=U''_{k_f}(0)^{1/2}$ in units of $k_0$
and the difference in free-energy density between the two vacua in units
of $m^3_f$.
Here $m_f^{-1}$ corresponds to the correlation length in the false
vacuum and $\Delta U/m^3_{k_f}$ can be related to observable quantities like
the jump in the order parameter or the latent heat if $\lx_{k_0}/k_0$ is
kept fixed.
(Similarly, for given $\lx_{k_f}/m_{k_f}$, we can relate
$\Delta U/m^3_{k_f}$  to $h$ in the approximation of eq.~(\ref{eq:two21})
using eq.~(\ref{deltaU})).
The spinodal line corresponds to the vertical axis, as for
$m_f=0$ the origin of the potential turns into a maximum.
The critical line corresponds to the horizontal axis.
The origin is the endpoint of the critical line.
During the evolution between the scales $k_0$ and $k_f$,
the quartic coupling becomes larger than
its initial value $\lx_{k_0}/k_0=0.1$.
All the potentials we have studied have an approximate form similar to
eq.~(\ref{eq:two21}) with $h\lta 0.9$.
From our discussion in the
previous section and fig.~2 we expect that $R$
is approximately given by eq.~(\ref{eq:appr})
with $T(h)\gta 0.3$ for $h\lta 0.9$.
This indicates that $R \gta 1$
for $m_f/k_0\lta  0.05$ even far from the spinodal line.
This expectation is confirmed by fig.~4.
Even for theories with a significant exponential suppression for
the estimated nucleation rate
we expect $R \sim 1$ near $m_f/k_0 \approx 0.05$.

\paragraph{6. Final comments:}
In the above discussion we have normalized the dimensionful quantities
such as $I$ and $A_k$
with respect to $k_f$, which is the natural scale
associated with tunnelling. In the context of thermal field theories,
the nucleation rate is often
expressed in units of the temperature, which may be identified with
the scale $k_0$ in our approach. As $k_f$ can be substantially lower
than $k_0$, the breakdown of the semiclassical approximation
(signalled by $R \sim 1$)
can occur for values of $I/k_0^4$ much smaller than those of $I/k_f^4$.

For a dynamical process with
temperature slowly varying on
a characteristic time scale $t_{\rm ch}$,
the phase transition
will essentially take place for $I/k_f^3=t^{-1}_{\rm ch}$.
For a known temperature dependence of the parameters in eq.~(\ref{eq:two20})
within the region of validity of Langer's theory, our
results permit a precise prediction of the amount of
supercooling by extracting from fig.~3 the effective transition
temperature. An approximate value of this temperature can also
be obtained by determining the bubble profile numerically and
using eq.~(\ref{eq:appr}) in order to estimate
the pre-exponential factor.

Finally, we point out that realistic statistical systems often
have large
dimensionless couplings $\lx_{k_0}/k_0\sim 10$.
Our results
indicate that Langer's homogeneous nucleation theory breaks down for
such systems even for small correlation lengths in the metastable
phase ($m_f/k_0 \sim 1$).
Furthermore, for a large enough correlation length the first-order phase
transition displays universal behaviour with large
$\lx_{k_f}/m_f\approx 5$, independently of the short-distance
couplings \cite{trans,B}. We conclude that a saddle-point approximation for
the fluctuations around the critical bubble does not give accrurate
results in the universal region.

\paragraph{Acknowledgements:}
The work of N.T. and C.W. is supported by the E.C. under TMR contract
Nos. ERBF\-MRX--CT96--0090 and ERBF-MRX-CT97-0122.


\end{document}